\documentstyle[preprint,aps]{revtex}
\begin{document}


\title{Band gap renormalization in photoexcited semiconductor quantum wire
structures in the {\it GW} approximation}
\author{E. H.\ Hwang and S. \ Das Sarma}
\address{Department of physics, University of Maryland, College Park,
Maryland  20742-4111 } 
\date{\today}
\maketitle

\begin{abstract}

We investigate the dynamical self-energy corrections of the
electron-hole plasma due to
electron-electron and electron-phonon interactions at the band edges
of a  quasi-one dimensional (1D) photoexcited electron-hole plasma.
The leading-order $GW$ dynamical screening approximation
is used in the calculation by treating
electron-electron Coulomb interaction and electron-optical phonon
Fr\"{o}hlich interaction on an equal footing. 
We calculate the exchange-correlation induced band gap renormalization
(BGR) as a function of the 
electron-hole plasma density and the quantum wire width. The
calculated BGR shows 
good agreement with existing experimental results, and 
the BGR normalized by the effective quasi-1D excitonic Rydberg
exhibits an approximate one-parameter universality.


\end{abstract}

\newpage

A highly dense electron-hole plasma (EHP) can be generated in a wide
variety of semiconductors by optical pumping.
The band structure and the optical properties of highly
excited semiconductors differ from those calculated for
noninteracting electron-hole pairs due to many-body
exchange-correlation 
effects arising from the EHP\cite{review,review1}. 
One of the important  many-body effects in high density EHP is a
density-dependent renormalization of the fundamental band gap of the
semiconductor, which
causes an increasing absorption in the spectral region below the lowest 
exciton resonance. 
The exchange-correlation correction of the fundamental band gap due
to the presence of 
free carriers (electrons in the conduction band and holes in the
valence band) in the system is referred to as the band gap
renormalization (BGR) 
effect. Optical nonlinearities, which are  strongly 
influenced by Coulomb interaction in the EHP, are typically
associated with the band-gap renormalization 
phenomenon. 
The band gap renormalization has been widely studied 
in bulk and quasi-two dimensional (quantum well) semiconductors
\cite{review,review1,dassarma}. 
In recent years, quasi-one dimensional  semiconductor quantum
wires (QW) have been 
fabricated with atomic scale definition, and QW optical
properties  have been studied
for their potential device applications such as semiconductor lasers.
There has, however, been little work on the BGR in QW systems, 
both experimentally \cite{cin} and
theoretically\cite{hu,tanatar}.  
So far most calculations have been done in the static screening
approximation or in the simple plasmon-pole
approximation, which is a simplified version of the random-phase
approximation  (RPA). 
The plasmon-pole approximation consists of ignoring the 
weight in the single particle excitations and assuming that all 
free carrier contributions  to the dynamical dielectric function
lies at the effective plasma frequency
$\omega_{p}$. 
The advantages of the plasmon-pole approximation are its
mathematical simplicity and simple physical meaning.
However, a certain degree of
arbitrariness in the choice of the effective plasmon pole parameters
which are needed to satisfy the $f$-sum rule and
the static Kramers-Kroning relation 
leads to considerable difficulties
in applying the theory to semiconductors with complex band structures.
In this paper, we calculate the BGR of the quantum wire structures
based on the RPA dynamical screening $GW$ scheme by taking into
account the full frequency 
dependent dielectric response in the {\it two component} one
dimensional (1D) EHP.

Most high quality QW structures of our interest are fabricated in 
weakly polar III-V semiconductors. 
In a polar semiconductor, free carriers couple to the
longitudinal optical (LO)-phonons of the underlying lattice through the
long-range polar Fr\"{o}hlich interaction.
The carrier-LO phonon interaction leads to polaronic 
many-body renormalization of the single particle free carrier
properties. Even though the weakly polar III-V materials have rather
small Fr\"{o}hlich coupling constants, the electronic properties can
still be substantially modified by the Fr\"{o}hlich interaction.
Thus, inclusion of dynamical electron-phonon
interaction in the theory is important since in the quantum well wires 
made of weakly polar materials Fr\"{o}hlich interaction produces
observable many-body corrections. 
In spite of substantial current interest in the properties of polar
quasi-one dimensional QW systems
existing in GaAs quantum wires, the full
many-body problem that includes dynamical screening and treats
electron-electron and electron-phonon interactions on equal footing
has not yet been worked out.

Our goal is to calculate the BGR of a coupled 1D
electron-phonon many-body system treating electrons and phonons on an
equal footing. 
There has been no detailed quantitative study of 1D quasiparticle 
properties including both 
electron--electron and electron--phonon interactions.
We provide such an analysis in this
article based on the leading-order many-body perturbation theory.
We calculate the self-energy corrections to the band edges (the
highest valence- and the lowest conduction-band edges) in the presence
of the EHP density $n_e = n_h$ of the electrons ($n_e$) and the
holes($n_h$) for quantum well wires of various thicknesses and for a
number of different semiconductor systems. 
To the best of our knowledge, this is the first calculation of
electronic many-body BGR correction in QW systems including full
effects of both the dynamical electron-electron and Fr\"{o}hlich
electron--LO-phonon  interactions treated on
an equal footing. 
We find that the calculated  BGR using the full RPA dielectric
function agrees very well with available experimental results
\cite{cin} and depends 
on both the electron-hole density and quantum wire widths. 
Our calculated BGR in quantum wires  shows an
approximate materials independent two-parameter universality as a
function of the {\it scaled} plasma density and wire
width.  We find that this two-parameter universality of the BGR can be
reduced to 
a single universal curve when the BGR scaled in the units of the
appropriate quasi-1D effective excitonic Rydberg is expressed as a
function of the dimensionless 1D electron-hole density  
parameter, $r_s=1/(2n a_B)$ with a plasma density $n$ and an effective
Bohr radius $a_B$.

We assume that our quasi-1D QW  system has an  infinite
square-well confinement with a finite width ($a$) in the 
y-direction  and zero thickness in the z-direction, which is one of
the simplest 1D confinement models.  Because of the universality
mentioned above we believe that our results are valid for more general
1D confinement situations. Only one kind of
electrons and holes with isotropic, parabolic dispersion in a direct
gap semiconductor is assumed to
exist, thus neglecting most of the band-structure complications of the
valence subbands. This should be an adequate approximation for
calculating the band edge BGR. We
consider here the $T=0K$  situation with only the 
lowest conduction subband (for the electrons) and the highest valence
subband (for the holes) occupied. The effective mass approximation is
expected to be fairly well-valid under the experimental conditions and
we will assume that uncritically for our theory.
Our main calculations use the $GW$ approximation for the
self-energy\cite{ando,hedin}. This is the leading term in an iterative
expansion of the 
self-energy operator in powers of the dynamically screened effective
electron-electron (including both Coulomb and Fr\"{o}hlich couplings)
interaction $W$, and has been shown to yield an 
excellent description of quasiparticle energies in semiconductors
\cite{hedin,ja,hwang}.

The BGR is given by the sum of the self-energies for electrons and
holes at band edges
\begin{equation}
\Delta = {\rm Re}\Sigma_e(0,0) + {\rm Re}\Sigma_h(0,0).
\end{equation}
The total electronic self-energy within the leading order effective
dynamical interaction in a two-component electron-hole plasma is
\begin{equation}
\Sigma_e(k,\omega) = i\int \frac{dq}{2\pi}
\int \frac{d\omega'}{2\pi} G_0(k-q,\omega-\omega')
\frac{V_{t}(q,\omega')} {\epsilon(q,\omega')},
\label{sigma}
\end{equation}
where $G_0$ is the Green's function for the noninteracting electron
gas,  $V_{t} = V_c + V_{\rm ph}$ the total effective interaction, and
$\epsilon(q,\omega) = 1-V_t(q,\omega) \Pi_0(q,\omega)$ the effective
dynamical dielectric function. Here $V_c$, $V_{\rm ph}$, and 
$\Pi_0(q,\omega)$ are the direct Coulomb interaction,  the
LO-phonon mediated electron--electron interaction, and  the
irreducible 1D noninteracting polarizability, respectively.
In our extreme quantum limit model
where only the lowest 1D subband is occupied by the electrons, we
obtain the interaction matrix elements by taking the quantizing
confinement potential to be of infinite square well type\cite{hu,hwang}
and the LO-phonon mediated electron--electron interaction is dependent
on wave vector and frequency, $V_{\rm ph}(q,\omega)=M_{\rm
q}^2D_0(\omega)$, where
$M_{\rm q}$ is the effective 1D Fr\"{o}hlich interaction matrix
element and 
$D_0(\omega)$ the unperturbed retarded bare LO-phonon propagator
\cite{hwang}. 
Since our system is a two-component system (electron-hole plasma), the
polarizability function is a sum of electron and hole
polarizabilities, $\Pi_0(q,\omega) = \Pi_{0e}(q,\omega) +
\Pi_{0h}(q,\omega)$. 
The formalism for holes, within our parabolic band approximation, is
the same as that for electrons and the only modification needed in
Eq. (\ref{sigma}) is the substitution of $m_e$ by $m_h$.  The
self-energy 
calculation is standard and more details can be found in the
literature \cite{hwang}.

In Fig. \ref{fig1} we show our calculated exchange-correlation-induced
band gap renormalization for quasi-1D quantum wires, which is scaled by
the effective 3D Rydberg, $E_0 = e^2/(2a_B\epsilon_0)$, as a function
of the effective electron-hole plasma density parameter, $r_s=1/(2n
a_B)$.
We also show in Fig. \ref{fig1} the band gap renormalization for
various  
quantum wire widths. The BGR increases with decreasing wire width.
The QW BGR depends rather strongly on the wire width because the 1D
Coulomb interaction matrix element is a sensitive function of the
width even in the long wavelength limit.
We have carried out the calculation of BGR as a function of
carrier density and wire width for a number of III-V semiconductor
materials. 
We find that, when expressed in suitable dimensionless units
(the effective 3D Rydberg and the effective 1D density parameter
$r_s$) as shown in Fig. \ref{fig1},  the band gap renormalization in
quasi-1D systems is an approximate two-parameter universal function,
{\it i.e.},  a universal function of the
effective density and the wire confinement width. (Since the lowest
exciton state is not well defined in the ideal zero width 
limit of the 1D electron-hole system \cite{1d}, we use the effective
3D Rydberg to scale the band gap renormalization.)
The inset in the Fig. \ref{fig1} shows the BGR for the quasi-1D
{\it GaAS-AlGaAs} system as a function of the electron-hole plasma
density $n$ together with the available
experimental data from Ref. \onlinecite{cin} , in which the BGR is
measured in confined 
quantum well structures with lateral widths of $600 \pm 50 \AA$. 
For GaAs we use the parameters $m_e=0.067m_0$, $m_h=0.45m_0$, and
the background dielectric constant $\epsilon_0 = 13$.
We find that the calculated  BGR using the full RPA dielectric
function agrees well with available experimental data \cite{cin}.
Better quantitative agreement between experiment and theory 
may require  a finite temperature 
calculation including higher subband filling effects,
which have been neglected in this calculation.

In bulk semiconductors the exchange-correlation
energy is essentially independent \cite{3d,dassarma} of band-structure
details when 
expressed in appropriate dimensionless units. This leads to a
universal form for the BGR when the self-energy corrections and the
carrier density are expressed in appropriate rescaled units of the
effective excitonic Rydberg and the normalized interparticle 
separation $r_s$, 
respectively \cite{3d}.  A similar universality holds for two
dimensional (2D) systems when the band gap is expressed in effective
2D Rydberg and the 
electron-hole density in the 2D $r_s$ parameter\cite{2d}. 
In quasi-2D quantum well systems, however, the BGR depends on the well
width and  is found to be a  two-parameter universal function of
the effective 2D $r_s$ parameter and well
width. It was shown \cite{dassarma}  that this
two-parameter universality can be reduced to an approximate
one-parameter universality by choosing suitable {\it quasi}-2D
Bohr radius and effective Rydberg as the effective length and energy
scaling units, respectively.
In order to investigate whether the two-parameter universality of
quasi-1D BGR in Fig. \ref{fig1}
can be reduced to a one-parameter universality by suitably rescaling
energy and length units, we calculate the quasi-1D exciton binding
energy, which would be used as a suitable energy rescaling unit.  The
exciton energy of the quasi-1D system has earlier been 
calculated in the cylindrical wire with a finite radius smaller than the
bulk exciton radius \cite{1dex}. 
However, in order to be consistent with
our BGR calculation we use the quasi-1D Coulomb interaction of the
quantum wire with a zero
thickness in $z$-direction and a finite width $a$ in $y$-direction to
calculate the 1D excitonic binding energy within our confinement
model. In this approximation and within the one subband extreme
quantum limit the effective 
quasi-1D Coulomb interaction is given by
\begin{equation}
V_{q1d}(x-x')=\int \int dydy' V_{2D}(x-x',y-y')|\phi(y)|^2|\phi(y')|^2,
\label{vq1d}
\end{equation}
where $V_{2D}(x,y)=e^2/(\epsilon_{0}\sqrt{x^2 + y^2})$ is the
2D Coulomb potential and $\phi(y)$ is the single particle wave function 
in the unperturbed quantum well wire with infinite barriers.
The Fourier transform of Eq. (\ref{vq1d}) is used as the bare Coulomb
interaction to calculate the
self-energy of the electron-hole plasma in Eq. (\ref{sigma}).
If we neglect the motion of carriers in y-direction,  the relative
motion of a quasi-1D electron-hole pair is described by 
the 1D Schr\"{o}dinger equation with the effective quasi-1D Coulomb
interaction given in Eq. (\ref{vq1d}). 
In Fig. \ref{fig2} we show our numerically calculated two lowest
excitonic binding 
energies $E_b(a)$ for a {\it GaAs-AlGaAs} quantum well wire as a
function of the 
width $a$ (the ground state $n=1$ and the first excited state $n=2$).
Comparing with the measured exciton energy ($E_b \approx 12$ meV) 
for a GaAs quantum wire with the lateral width $a = 600 \pm 50$
$\AA$\cite{cin1}, our calculation gives reasonable agreement.
Using the calculated  effective exciton binding energy
(Fig. \ref{fig2}) we rescale the calculated QW band gap
renormalization to investigate the applicability of a one-parameter QW 
BGR universality. In Fig. \ref{fig3}, we show the rescaled BGR in the
unit of the 
effective binding energy of the exciton as a function of the density
parameter $r_s$. As one can see from Fig. \ref{fig3}, this rescaling of
units in terms of {\it quasi}-1D effective parameters produces an
approximate one-parameter universality.
(Note that we use the 3D Bohr radius $a_B$ to rescale the lengths.)
In order to check the general applicability of our universal BGR
results, experimental data on different III-V 
quasi-1D semiconductor QW systems are needed. It would be interesting
to verify our predicted one parameter BGR universality by measuring
the BGR in different III-V materials for QW of different widths.

In conclusion, we investigate the band-gap renormalization of quasi-1D
electron hole plasma 
systems in semiconductor QW structures including full dynamical
effects of both the electron-electron 
and electron-phonon interactions
treating electrons and phonons on an equal footing. 
Our calculated BGR agrees well with the available experimental data.
We derive an approximate
one parameter universality in the band gap renormalization (rescaled
by the effective quasi-1D excitonic Rydberg) as a
function of the effective 1D density parameter $r_s$, which should be
experimentally checked.

\section*{ACKNOWLEDGMENTS}
This work is supported by the U.S.-ARO and the U.S.-ONR.

\begin{figure}
\caption{The calculated band-gap remormalization $\Delta$ as a
function of the 
effective electron-hole pair density parameter $r_s = 1/(2n
a_B)$ for different quantum wire widths, where $a_B$ is the bulk
effective Bohr radius. 
The BGR is scaled by the 3D exciton binding energy $E_0 =
e^2/(2a_B\epsilon_0)$. 
In the inset we show the BGR for the quasi-1D {\it GaAS-AlGaAs}
system as a function of the electron-hole plasma density $n$.
The experimental points are taken from Ref. [4]. The middle
thick line is for confinement width $a=600$ $\AA$ and the upper
(lower) thin line for $a=650$ $\AA$ ($a=550$ $\AA$). 
}
\label{fig1}
\end{figure}

\begin{figure}
\caption{The two lowest effective exciton binding energies of a quasi-1D
GaAs QW system as a function of
the confinement width $a$. $n=1$ ($n=2$) is for the ground (the first
excited) state.
}
\label{fig2}
\end{figure}

\begin{figure}
\caption{The approximate one-parameter universality of the band-gap
renormalization of the quasi-1D system as a function of the effective
density parameter $r_s$.
We rescale the BGR in Fig. \ref{fig1} in the unit of the
effective quasi-1D exciton binding energy of Fig. \ref{fig2}. The
experimental points from ref. [4] are also shown.
}
\label{fig3}
\end{figure}

\end{document}